\documentclass[12pt]{article}
\usepackage{vmargin}
\setpapersize{A4}
\setmarginsrb{2cm}{3cm}{2cm}{2cm}{0pt}{0pt}{12pt}{24pt}
\usepackage[bf,sf,medium]{titlesec}
\usepackage{cite} 
\usepackage{graphicx}
\usepackage[labelfont=bf]{caption}
\usepackage{pifont}
\newcommand{\xmark}{\ding{55}}
\usepackage{authblk}
\usepackage{enumitem}
\usepackage[labelformat=simple]{subcaption}
\usepackage{floatpag}
\usepackage{amsmath}
\usepackage{amssymb}
\usepackage{algorithm}
\usepackage[noend]{algpseudocode}

\newtheorem{proposition}{Proposition}
\usepackage{relsize}
\usepackage{xspace}

\usepackage{chngpage}
\usepackage{longtable}
\usepackage[breaklinks]{hyperref}
\hypersetup{
	colorlinks=true,
	linkcolor=blue,
	citecolor=blue
}
\DeclareMathOperator{\Expectation}{E}

\usepackage{booktabs}
\usepackage{tabularx}

\newcommand{\tabhead}[1]{\textbf{#1}}

\usepackage[titletoc]{appendix}

\floatpagestyle{empty}
\usepackage{bm}
\usepackage{breakurl}
\usepackage{times}
\usepackage{setspace}
\algnewcommand{\aand}{\textbf{and}\xspace}
\algnewcommand{\oor}{\textbf{or}\xspace}
\algnewcommand\True{\textbf{true},\space}
\algnewcommand\false{\textbf{false}\space}
\algnewcommand\False{\textbf{false},\space}
\algnewcommand\Print{\State \texttt{Print: }}
\algnewcommand\Raise{\State \texttt{raise exception: }}
\algdef{SE}[DOWHILE]{Do}{doWhile}{\algorithmicdo}[1]{\algorithmicwhile\ #1}

\begin{document}

\title{A Generalized Population Dynamics Model of a City and an Algorithm for Engineering Regime Shifts}
\author[1,2]{James P.L. Tan}
\affil[1]{Interdisciplinary Graduate School, Nanyang Technological University, Singapore}
\affil[2]{Complexity Institute, Nanyang Technological University, Singapore}
\date{}
\maketitle

\begin{abstract}
Measures of wealth and production have been found to scale superlinearly with the population of a city. Therefore, it makes economic sense for humans to congregate together in dense settlements. A recent model of population dynamics showed that population growth can become superexponential due to the superlinear scaling of production with population in a city. Here, we generalize this population dynamics model and demonstrate the existence of multiple stable equilibrium points, showing how population growth can be stymied by a poor economic environment. This occurs when the goods and services produced by the city become less profitable due to a lack of diversification in the city's economy. Then, relying on critical slowing down signals related to the stability of an equilibrium point, we present an algorithm for engineering regime shifts such that a city at a stable equilibrium point may continue to grow again. The generality of the model and the algorithm used here implies that the model and algorithm need not be restricted to urban systems; they are easily applicable to other types of systems where the assumptions used are valid. 
\end{abstract}

Cities are large and dense spatial agglomerations of humans and their socioeconomic activities. The growth of cities results in distinct spatial patterns of settlement and human activity that have been the subject of extensive research over the past decades. There is now a growing consensus that the processes that give rise to urban spatial patterns are localized, resulting in urban growth driven from the bottom up \cite{BattyScience1, BattyNature1}. There is also a considerable amount of research effort focused on the morphology of urban growth where form is more emphasized rather than function \cite{BattyEnvironment, MakseNature1, Louf1, Fujita1}. In urban economics, a few well-known results exist concerning the optimal town size \cite{Mirrlees,Dixit}. However, these studies are usually more concerned on the spatial rather than the temporal aspect of urban growth. To obtain new insights into the evolution of a city and why some cities thrive where other cities fail, consideration must also be given to the  temporal aspect of urban growth and the factors that drive this growth. Examples of work that has been done in this area are the modeling of retail and residential spaces of a city using difference equations by  Beaumont, Clarke and Wilson \cite{Beaumont}, and the modeling of population migration within a city by Weidlich and Haag \cite{Weidlich}. In this paper, we build upon previous work by Bettencourt et al. and consider a simple population dynamics model driven by the population migration that may occur to take advantage of newly emerged economic opportunities \cite{BettencourtPNAS}. 

Cities represent places of economic opportunity for the population migration of humans. Individuals and corporations come together for the exchange of goods and services in close proximity \cite{Glaeser1,Dumais1,Diamond1,Glaeser2,Puga1}. Indeed, empirical data from cities indicate that measures of wealth and production scale superlinearly with the population of a city \cite{BettencourtPNAS}. The scaling appears in the form of power laws $Y= Y_0 N^\beta$ \cite{BettencourtPNAS} where $Y$ is a property of the city, $Y_0$ is a constant, $N$ is the population of the city and $\beta$ is the scaling exponent. Superlinear scaling occurs when $\beta>1$ and sublinear scaling occurs when $\beta<1$. There has been some controversy surrounding this result because it has been observed that the scaling exponent varies quite sensitively to the definition of a city's boundaries over which properties of a city like wealth and production are aggregated \cite{Arcaute}. However, a consistent scaling exponent can be observed across multiple cities if the definition of a city's boundaries is able to capture urban functionality \cite{Bettencourt,Masucci}. Furthermore, the empirical exponents can be theoretically predicted by considering the social interactions of its residents on a spatial network \cite{BettencourtScience1, PanNatComm1}. For properties related to production and growth, $\beta$ is theoretically estimated at $\beta\approx 1.17$. While properties related to production and growth might scale superlinearly with the population, undesirable properties like crime can also scale superlinearly with city size \cite{Cullen, Gould,Glaeser,BettencourtPNAS,Alves}. These are obvious trade-offs that economic migrants must make when choosing to settle in a city. Therefore, a city will not grow if the disadvantages that come with agglomeration outweighs the advantages that come with it. Clearly, this has not been the case for all cities especially with the ongoing process of urbanization in the modern world \cite{Grimm}. With regard to production and wealth, it is generally more economically viable for a population to congregate and settle in a city as economic output increases superlinearly with population. However, there are many examples of cities initially prospering and then failing economically, stagnating and even undergoing urban decay. The city of Detroit in the United States is one such example. From 1900 to 1950, Detroit's population increased roughly six times from 285,704 to 1,849,568 before starting a sustained decline to 713,777 by 2010 (Fig. \ref{fig:PCH}(b)). The rise of Detroit in the first half of the twentieth century is attributed to its automobile manufacturing industry, with the automobile industry of the United States consolidating and agglomerating around Detroit \cite{Klepper1, Hyde1}. While there are many hypothesized reasons for Detroit's decline in the second half of the twentieth century, they all share a common theme of deindustralization of the automobile industry in Detroit as the city became less attractive to automobile manufacturers \cite{Sugrue1,Sugrue2}. Detroit's over-reliance on the automobile industry and its failure to properly diversify into other profitable industries led to an economic vacuum as automobile manufacturers left the city, driving a population decline amidst a lack of jobs. Therefore, even though wealth and economic output increases superlinearly with the population of a city, any population growth from a growing economy must also be contingent on the profitability of the city's industrial output among other socioeconomic factors.  

As population growth becomes stymied due to economic factors, the population might languish in a stable population regime. Any small perturbation to the population in such a stable regime will only decay with time. Hence, it is important to be able to control and engineer a regime shift out of this stable regime so that the population may grow again. Regime shifts are discontinuous in the sense that they can involve large changes to a state variable in a short amount of time. The literature on regime shifts is mostly concerned with how to avoid rather than to control them \cite{Martin}. This is because regime shifts are mostly negatively associated with unwanted phenomena like the desertification of vegetation covered regions or wildlife population collapse \cite{Hirota1, Hutchings1}. However, if one is confident of the direction of a regime shift, then a regime shift can become beneficial. In this paper, we will first present a generalized model of urban population growth driven by population migration due to economic opportunity. Then we will show that it is possible for growth to be stymied with the presence of multiple stable equilibrium points in the population. Finally, we will outline and demonstrate a generic algorithm to engineer regime shifts out of these equilibrium points such that the population may grow again. 

\section*{Generalized one-dimensional model of population growth}
A model of urban population growth by Bettencourt et al. is
\begin{align} \label{eq:Bettencourtmodel}
\frac{ \mathrm dN}{\mathrm dt} = \frac{Y_0}{E} N^\beta - \frac{R}{E} N^\alpha, 
\end{align}
where $E$ is the resources needed to add an individual to the city per unit time, $Y_0N^\beta$ is the resources generated by the city per unit time, $RN^\alpha$ is the resources consumed by the city per unit time, and $\alpha$ and $\beta$ are scaling exponents \cite{BettencourtPNAS}. Therefore, this model assumes that the surplus resources generated by the city ($Y_0 N^\beta - RN^\alpha$) goes towards growing the population. More specifically, this can happen when the extra wealth or resources generates more demand for goods and services, creating jobs and economic opportunities for migrants. Depending on the initial conditions and the exponents, population growth can be growing or decaying towards a carrying capacity, collapsing, increasing superexponentially or increasing exponentially. For cities in the face of unimpeded growth and a linear consumption of resources, we expect $\beta=1.17$ and $\alpha=1$. In this case, $(R/Y_0)^{1/(\beta-1)}$ is an equilibrium point of the system. This leads to superexponential growth when $N(0) > (R/Y_0)^{1/(\beta-1)}$ and population collapse when $N(0) < (R/Y_0)^{1/(\beta-1)}$ \cite{BettencourtPNAS}. Superexponential growth of a city is plausible with the process of urbanization as the rural population migrates to the city. However, population growth will eventually become biologically limited to exponential growth in the absence of population migration. Additionally, population growth can be stymied or even reversed from any number of different factors like natural disasters, foreign invasions, changes in social trends, and even ineffectual urban planning \cite{Jacobs1}. Here, we shall consider economic reasons for the population stagnation or decline of cities by generalizing the resource production and consumption rate of cities. 

By generalizing Eq. \ref{eq:Bettencourtmodel}, we obtain a simple one-dimensional model of urban population growth which is
\begin{align} \label{eq:model}
\frac{ \mathrm dN}{\mathrm dt} = f(N) = P(N) - C(N), 
\end{align}
where $N\geq 0$, $P(N)$ is the rate of production of resources and wealth in the city and $C(N)$ is the rate of consumption of the resources and wealth in the city, including the costs that come with agglomeration in the city. Thus, we require $P(N)$ and $C(N)$ to be non-negative and strictly increasing functions of $N$. In the context of this generalized model, we define superlinear and sublinear scaling using the second derivative; a function of the population $g(N)$ scales superlinearly with the population at $N_0$ if $g''(N_0)>0$, sublinearly with the population at $N_0$ if $g''(N_0)<0$, and linearly with the population at $N_0$ if $g''(N_0)=0$ provided the second derivative of $g(N)$ exists at $N_0$. Let $P(N)$ and $C(N)$ be continuous functions that intersect each other $n$ times resulting in $n$ intersection points, where $n$ is a positive integer. The intersection points of the functions $P(N)$ and $C(N)$ give the equilibrium points of the dynamical system. Let $N^*=(N^*_1, N^*_2, N^*_3, \dots, N^*_n)$ represent the sequence of equilibrium point solutions to the dynamical system in increasing order i.e. $P(N^*_i)-C(N^*_i)=0$ for any $i \in \{1,2,\dots,n\}$. Here, $N^*_1=0$ since there has to be an equilibrium point at the origin for a population growth model. $N^*_1$ is stable if $f(N^*_{1+})<0$ and unstable if $f(N^*_{1+})>0$. $N^*_{1+}$ is any number that satisfies $N^*_1 < N^*_{1+} < N^*_2$. For $i\neq 1$, the equilibrium point $N^*_i$ is stable if $f(N^*_{i+})<0$ and $f(N^*_{i-})>0$. Here, $N^*_{i+}$ is any number that satisfies $N^*_i<N^*_{i+}<N^*_{i+1}$ if $i \neq n$ or $N^*_i<N^*_{i+}$ if $i=n$. $N^*_{i-}$ is any number that satisfies $N^*_{i-1}<N^*_{i-}<N^*_i$. Conversely, $N^*_i$ is unstable if $f(N^*_{i+})>0$ and $f(N^*_{i-})<0$. $N^*_i$ is half-stable if $f(N^*_{i+})$ and $f(N^*_{i-})$ have the same signs. Stable equilibrium points are known as regimes and transitions between regimes are known as regime shifts or critical transitions. An obvious but interesting result from this generalized model is that the stability of the equilibrium points always alternate between stable and unstable, not counting the half-stable equilibrium points (Proposition \ref{proposition:alternatingstability} in the Appendix). 

\begin{figure*}[!ht]
\centering
\centerline{\includegraphics[scale=0.45]{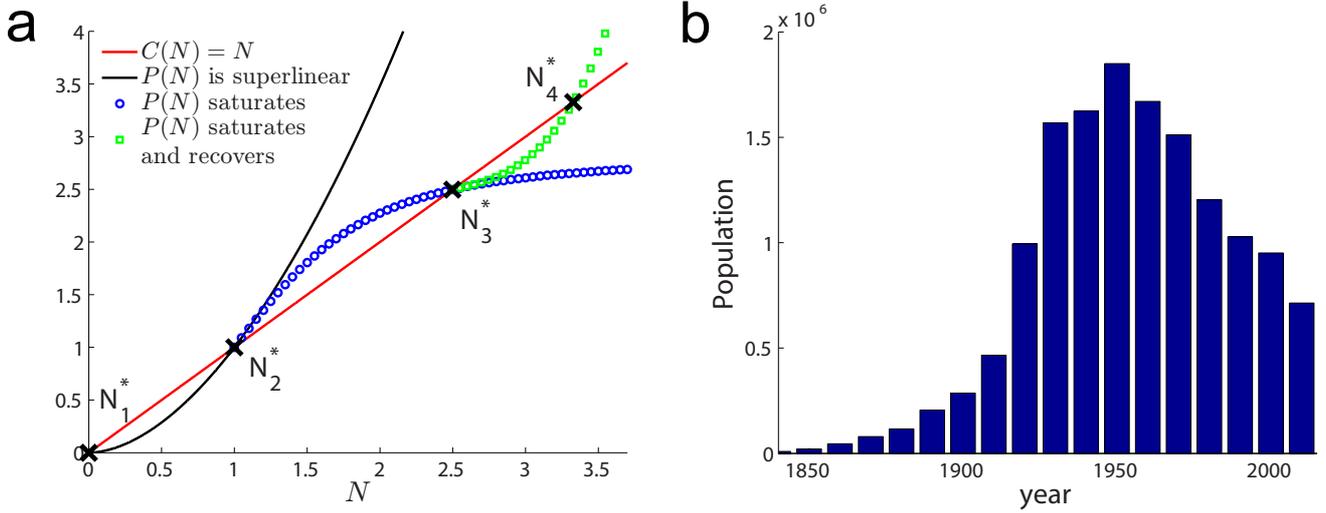}}
\caption{(a) $P(N)$ and $C(N)$ for a city with a consumption that scales linearly with population. The equilibrium points given by the intersections of $P(N)$ and $C(N)$ are marked by crosses on the plot. $P(N)$ is shown to scale superlinearly with the population (black line). After the second equilibrium point $N^*_2$ which is unstable, an alternate scenario where $P(N)$ begins to saturate is shown by the blue circles. After the third equilibrium point which is stable, another alternate scenario is presented where $P(N)$ begins to scale superlinearly again due to a diversification into profitable goods and services as shown by the green squares. (b) The population of Detroit by decade (Source: US Census Bureau \cite{Census1}). } 
\label{fig:PCH}
\end{figure*}

We now use this generalized model of population growth to explain economic obstacles to population growth in a city. In Bettencourt et al.'s model, $N^*_1$ is stable since $P(N)=Y_0N^{1.17} /E$ and $C(N)=RN /E$. The next and only other equilibrium point, $N^*_2$ is unstable. It should be noted that depending on $R$ and $Y_0$, it is possible in this model for $0<N^*_2<1$. This is simply the case when a population of one produces more than he consumes. When $N>N^*_2$, growth is superexponential (Fig. \ref{fig:PCH}(a)). Here, growth is unsustainable as the population approaches a singularity in finite time \cite{BettencourtPNAS}. In reality, we expect factors like the competition for resources to force $P(N)$ to grow sublinearly and eventually saturate with the population such that a superexponential growth ceases to perpetuate. In the case where a city's industry is not diversified, an abundance in a city's production output can also dent the growth of $P(N)$. With undiversified growth, $P(N)$ would intially scale superlinearly with $N$ when $N$ is slightly larger than $N^*_2$. In this growth phase, the city's population grows superexponentially. However, due to a lack of diversity, a glut of the city's products in the national or international market in the presence of a lack of demand will dent the growth of $P(N)$ when $N$ is substantially larger than $N^*_2$ so that $P(N)$ eventually grows sublinearly with $N$ and saturates at large $N$. Therefore, we expect a third equilibrium point $N^*_3$ which is stable in the dynamical system as $P(N)$ intersects $C(N)$ again from the top (Fig. \ref{fig:PCH}(a)). Hence, population growth in the city would cease at $N^*_3$. The ability of the city to grow again would then depend on whether it can diversify into other profitable industries or ramp up the profitability of its products. In the case of diversification into profitable industries, we model $P(N)$ to start scaling superlinearly again after $N^*_3$ due to potential diversification into profitable industries after $N^*_3$. This recovery of $P(N)$ after $N^*_3$ represents a conscious decision by city planners to foster growth, investment, and diversification into more profitable industries. Note that diversification into a profitable industry can be possible before $N^*_3$ so long as a city has the necessary population and environment to support it. But because we are modeling initially undiversified growth, we are modeling $P(N)$ such that the city does not diversify into profitable industries before $N^*_3$ possibly due to a lack of foresight or future planning. The production function would not be exact for every city that stagnates and recovers from initially undiversified growth, but this saturation and recovery are similar features that we model in $P(N)$ across these cities. At a certain point after $N^*_3$, a fourth unstable equilibrium point $N^*_4$ must be overcome such that $N$ is again in the growth phase i.e. $N>N^*_4$ (Fig. \ref{fig:PCH}(a)). To engineer such a regime shift out of $N^*_3$, investments must be made to attract a large enough population into the desired industry in the city within a short time so that $N>N^*_4$. If this new influx of population is not large enough such that the population is still in the basin of attraction of $N^*_3$ i.e. $N^*_3<N<N^*_4$, then the population would decay back towards $N^*_3$. We note that in this model, it is not necessary that $C(N)$ scales linearly with the population. For $N^*_3$ to exist, it is enough that $C(N)$ does not asymptotically approach any consumption level smaller than the saturation level of $P(N)$ after $N^*_2$. For $N^*_4$ to exist, $P(N)$ simply has to increase faster than $C(N)$ after $N^*_3$ such that $P(N)$ again intersects $C(N)$ from the bottom. 

In the context of the model presented, the population collapse in Detroit after the 1950s could be due to a regime shift to an equilibrium point with a lower population after approaching a bifurcation caused by the worsening economic situation in Detroit. In this case, $P(N)$ is also a function of profitability $p$ unrelated to diversification i.e. $P(N,p)$. As the profitability $p$ decreases, $P(N,p)$ would be strictly decreasing with decreasing $p$ for all $N$. Hence, the stable equilibrium point that Detroit was residing in would collide with an earlier unstable equilibrium point as $P(N,p)$ moves below $C(N)$, leading to a bifurcation and population collapse. This urban decline can also be seen in other cities that fail to reinvent and diversify their economies. For example, the city of Youngstown in Ohio, US experienced a population decline of about 60\% from its peak population in the 1960s. This is thought to be largely a result of an over-dependence on its steel industry, which collapsed leading to job losses and unemployment \cite{Fuechtmann, Safford, Rhodes}. Baltimore is another city in the US experiencing urban decline due to deindustrialization \cite{Berger}. More successful cities like London that have avoided urban decline despite deindustrialization have managed to grow by diversifying into the service industry \cite{Graham}. Taken to the extreme, urban decline can also be seen in the many abandoned mining towns of yore. When the dominant economic activity, in this case mining, ceased to be viable due to a depletion of resources, these towns were abandoned due to a lack of economic opportunity. Hence, economic opportunity is a necessary condition for a city to thrive and grow, as is also evident from the empirical scaling discovered by Bettencourt et al. and qualitative observations of economic diversity and urban decline by Jacobs \cite{Jacobs}. Before such an urban decline, the model of population growth presented here predicts the presence of stable equilibrium points which represents the stagnation of the population of a city. This presents an opportunity to engineer a regime shift out of the stable equilibrium point so that the population may grow again. 

As alluded to earlier, population and investment are needed in a new industry in order to overcome the basin of the stable equilibrium point. Investment in areas such as infrastructure, logistics, and land might be necessary so that the new industry can operate in the city. With the operational needs of a future industry taken care of, the next task is to attract companies and jobs, and with it, a large enough working population into the new industry so that this diversification endeavor is profitable i.e. overcoming the basin of attraction of the stable equilibrium point. The threshold in population needed for profitability is, as mentioned, due to the superlinear scaling effect of production \cite{BettencourtPNAS}. This scaling phenomenon stems from the effects of agglomeration in economic activity \cite{BettencourtScience1}. It is entirely possible that a population of one in the new industry produces more than he consumes at the outset which is effectively saying that the basin of attraction is less than one person and that the industry can grow and is profitable from one person. But we do not consider such a situation because it is a trivial affair to grow the industry once the infrastructure needed to support it is in place. For a larger basin of attraction, there is a need to attract companies and jobs so that it might prove necessary on the part of the city to subsidize the cost of setting up business in the city. An example is the rapid industrialization of Singapore in the second half of the 20th century through the efforts of the Economic Development Board of Singapore, which aggressively pursued policies that included subsidies to attract industries it deemed beneficial to the economic development of Singapore \cite{Schein,Siddiqui}. Of course, subsidizing the cost of setting up business might entail additional investments on the part of the city which might be risky if the basin of attraction is large. This is because the population that was attracted by such an investment to the city may not be large enough to overcome the basin of attraction, leading to a decay of the population back to the stable equilibrium point and squandering the investment made by the city in attracting companies and jobs into the city. Mathematically, in order to reduce the size of the basin, bifurcation parameters of the dynamical system can be altered such that the stable equilibrium point becomes destabilized. In this way, the basin of attraction becomes smaller and the equilibrium point approaches a bifurcation that annihilates the initial regime that the system was residing in. 

While the set of profitable industries is likely to be unique for each city, the forms of support and incentives a municipal government can offer are largely the same e.g. reduced corporate tax rates, land concessions, etc. These are probable bifurcation parameters because they can be tuned to increase the profitability of an industry so that $P(N)$ can be made to increase faster out of an equilibrium point, decreasing the basin of attraction and bringing the stable equilibrium point closer to a bifurcation. Hence, these support and incentives for cultivating and diversifying into a profitable industry could be potential bifurcation parameters that will work with the algorithm to engineer a successful regime shift in a stagnating city. It should be noted that accurate identification of the bifurcation parameters is not necessary for the algorithm that we present in this paper because the algorithm can ascertain whether a parameter can bring about the desired loss of stability when tuned. 

By destabilizing the stable equilibrium point first with a bifurcation parameter, we need not risk the investment not being large enough such that the population still resides in the basin of attraction of the initial regime after the investment, leading to a decay back towards the equilibrium point. However, there are two main problems that have to be addressed when trying to alter a parameter to bring about a bifurcation: (1) identifying the bifurcation parameter, and (2) determining what direction the resulting regime lies in after a bifurcation has occurred. Both of these problems can be solved by measuring critical slowing down (CSD) signals in the system. 

Critical slowing down signals are statistical signals arising from the phenomenon of critical slowing down, where the decay rate of perturbations to a dynamical system residing in an attractor becomes slower as the attractor approaches a bifurcation and loses stability \cite{SchefferScience1, SchefferNature1}. These signals have been detected in a wide variety of physical, natural and socioeconomic systems on the verge of undergoing critical transitions and regime shifts \cite{CarpenterScience1, Leemput1, Krkosek1, SchefferScience1}. By measuring these signals, we can tell whether a not a system is losing stability and approaching a bifurcation point. The skewness of fluctuations, itself an early warning signal to regime shifts \cite{GuttalEcolLett1}, also tells us the direction of regime shifts after some bifurcations. Because$ f(N)$ is continuous, the resulting regime (if it exists) will lie in the direction where the skewness is changing (positive for increasing skewness and negative for decreasing skewness). Fluctuations do not become skewed before a pitchfork bifurcation because the equilibrium point is symmetrically annihilated by unstable equilibrium points from both directions. However, if we approach the pitchfork bifurcation in the symmetry broken state, fluctuations do become skewed. Skewness as a direction of regime shift can also work for other types of bifurcations like the saddle-node bifurcation where there is an increasing lack of symmetry in the stability of the equilibrium point as a bifurcation point is approached. In the next section, we will go into detail on the decay rate phenomena that can be observed for the various local bifurcations. 

\section*{Decay rate phenomena in various local bifurcations} \label{sec:decayratephenomena}

The decay rate of a perturbation from any stable equilibrium point $x^*$ for a continuous one-dimensional dynamical system $\dot{x}=f(x)$ is governed by $f(x)$. Let $f(x)$ be a smooth function of $x$. The decay rates are symmetrical between both directions for the dynamical system residing at $x^*$ if $f(x)$ is an odd function of $x$ about $x^*$ in the basin of attraction of $x^*$ i.e. $f(x^*+\varepsilon)=-f(x^*-\varepsilon)$, where $\varepsilon$ is the magnitude of a perturbation from $x^*$ and is also any positive real number such that $x^*\pm \varepsilon$ is within the basin of attraction of $x^*$. Specifically, we define the symmetry of decay rates between both directions to be,
\begin{align}
S(\varepsilon)=\frac{\min \left( |g(\varepsilon)|, |g(-\varepsilon)|\right)}{\max \left( |g(\varepsilon)|, |g(-\varepsilon)|\right)}, 
\end{align}
where $g(y) = f(y) = f(x-x^*)$. Therefore, the decay rates are symmetric if $S(\varepsilon)=1$ and asymmetric if $0\leq S(\varepsilon)< 1$. 

\paragraph{Saddle-node bifurcations} The normal form of a saddle-node bifurcation is $f(x) = r+x^2$, where $r$ is the bifurcation parameter. If $r<0$, then $x^*_\pm = \pm \sqrt{-r}$ are equilibrium points, with $x^*_-$ being stable and $x^*_+$ being unstable. A saddle-node bifurcation occurs when $r$ is increased past zero which results in the annihilation of $x^*_-$ and $x^*_+$. The system is then propelled in the positive direction in the ensuing regime shift. By a translation in coordinates $y=x+\sqrt{-r}$ so that $\dot{y}=g(y)=y^2-2y\sqrt{-r}$, we see that as $r$ is increased towards zero, the decay rates become slower as $|g(\pm \varepsilon)|$ decreases. Furthermore, the decay rate is faster along the negative direction than the positive direction because $|g(\varepsilon)|<|g(-\varepsilon)|$. Therefore, the direction of the regime shift is the same as the direction with the weaker decay rate. The symmetry of decay rates between both directions is, 
\begin{align}
S(\varepsilon)=\frac{-\varepsilon^2+2\varepsilon\sqrt{-r}}{\varepsilon^2+2\varepsilon\sqrt{-r}}. 
\end{align}
Hence, we expect $S(\varepsilon)$ to decrease as $r$ is increased towards the saddle-node bifurcation. 

\paragraph{Transcritical bifurcations} The normal form of a transcritical bifurcation is $f(x) = rx-x^2$. The equilibrium points are $x^*_r = r$ and $x^*_0 = 0$. Without loss of generality, we consider the case when $r<0$. The equilibrium point $x^*_0$ is stable and the equilibrium point $x^*_r$ is unstable. As $r$ is increased past zero, a transcritical bifurcation occurs and the two equilibrium points swap stability. The system is then propelled in the negative direction in the ensuing regime shift if noise is present in the system. When $r$ is increased towards zero approaching the transcritical bifurcation, perturbations to the system from $x^*_0$ experience a decreasing decay rate as $|f(\pm\varepsilon)|$ decreases. Furthermore, $|f(\varepsilon)|>|f(-\varepsilon)|$ so that perturbations along the negative direction experience a slower decay rate than the positive direction. Therefore, the direction of the regime shift is the same as the direction with the weaker decay rate. The symmetry of decay rates between both directions is,
\begin{align}
S(\varepsilon) = \frac{-r\varepsilon-\varepsilon^2}{-r\varepsilon+\varepsilon^2}. 
\end{align}
Hence, we expect $S(\varepsilon)$ to decrease as $r$ is increased towards the transcritical bifurcation. 

\paragraph{Supercritical pitchfork bifurcations} The normal form of a supercritical pitchfork bifurcation is $f(x) = rx-x^3$. The equilibrium points are $x^*_0 = 0$ and $x^*_\pm= \pm\sqrt{r}$. The equilibrium point $x^*_0$ is stable when $r<0$ and unstable when $r>0$. The equilibrium points $x^*_\pm$ are stable when $r>0$ and do not exist when $r<0$. When $r<0$ and is increased past zero, a supercritical pitchfork bifurcation occurs where $x^*_0$ becomes unstable with the appearance of the two stable equilibrium points $x^*_{\pm}$ at $x^*$ resulting in no regime shifts. Let $\varepsilon_0$ be the magnitude of a perturbation from $x^*_0$. When $r<0$ and is increased towards zero, then we see that the decay rate decreases as $|f(\pm\varepsilon_0)|$ decreases. Furthermore, $f(x)$ is an odd function about $x^*_0$ so that $S(\varepsilon_0)=1$. Hence, the decay rates are symmetrical about $x^*_0$. When $r>0$ and is decreased past zero where the supercritical pitchfork bifurcation occurs, the two stable equilibrium points $x^*_\pm$ are annihilated. Without loss of generality, we consider the case of the equilibrium point $x^*_+$ approaching the bifurcation. Let $\varepsilon_+$ be a perturbation from $x^*_+$. By a translation $y=x - \sqrt{r}$ so that $g(y)=-2ry - 3\sqrt{r}y^2-y^3$, we see that as $r$ is decreased towards zero, the decay rate decreases as $|g(\pm \varepsilon_+)|$ decreases. Furthermore, $|g(\varepsilon_+)|>|g(-\varepsilon_+)|$ so that perturbations along the negative direction experience a slower decay rate than the positive direction. The symmetry of decay rates between both directions is,
\begin{align}
S(\varepsilon_+) = \frac{2r\varepsilon_+ - 3\sqrt{r}{\varepsilon_+}^2+{\varepsilon_+}^3}{2r\varepsilon_++3\sqrt{r}{\varepsilon_+}^2 +{\varepsilon_+}^3}
\end{align}
Hence, we expect $S(\varepsilon_+)$ to decrease as $r$ is decreased towards the supercritical pitchfork bifurcation. 

\paragraph{Subcritical pitchfork bifurcations} The normal form of a subcritical pitchfork bifurcation is $f(x) = rx+x^3$. The equilibrium points are $x^*_0=0$ and $x^*_\pm = \pm\sqrt{-r}$. When $r<0$, $x^*_0$ is stable and $x^*_\pm$ are unstable. When $r>0$, $x^*_0$ is unstable and $x^*_\pm$ do not exist. Hence, when $r<0$ and $r$ is increased past zero, a subcritical pitchfork bifurcation occurs such that $x^*_0$ becomes unstable with the appearance of two unstable equilibrium points $x^*_\pm$ at $x^*_0$. When $r<0$ and is increased towards zero, we see that the decay rate of perturbations decreases as $|f(\pm \varepsilon)|$ decreases. Furthermore, $f(x)$ is an odd function about $x^*_0$ so that $S(\varepsilon)=1$. Hence, the decay rates are symmetrical about $x^*_0$. 

\begin{table}[h!] 
\caption{Summary of phenomena for stable equilibrium points approaching local bifurcations in a one-dimensional dynamical system $\dot{x}=f(x)$. Here, $x^*_0=0$ and $x^*_{\pm}=\pm \sqrt{r}$ are stable equilibrium points in the normal form of a supercritical pitchfork bifurcation and $r$ is the bifurcation parameter.  Checkmarks and crossmarks refer to the presence and absence of the observable phenomena listed respectively. }
\label{table:CSD}
\centering
\begin{tabularx}{0.82\textwidth}{l  l  p{20mm}  p{35mm}}
\toprule
\tabhead{Bifurcation type} & \tabhead{Regime shift} & \tabhead{Decreasing decay rate} & \tabhead{Decreasing decay rate symmetry}\\
 \midrule
Saddle-node & \checkmark &  \checkmark & \checkmark \\ 
Transcritical & \checkmark & \checkmark & \checkmark \\ 
Supercritical pitchfork, $x^*_0$ & \xmark & \checkmark & \xmark  \\ 
Supercritical pitchfork, $x^*_{\pm}$ & \xmark & \checkmark & \checkmark  \\ 
Subcritical pitchfork & \checkmark & \checkmark & \xmark \\
\bottomrule
\end{tabularx}
\end{table}

\section*{Manipulating regime shifts}
Based on the possible decay rate phenomena that can be observed for various bifurcations (Table \ref{table:CSD}), we present an algorithm to manipulate regime shifts in Eq. \ref{eq:model} by inducing bifurcations so that the system may escape regimes to increase or decrease the value of $N$ at equilibrium. This algorithm relies on the results of the previous section which stipulates that the direction of regime shifts is the same as the direction of weaker decay rates of perturbations, if asymmetry of the decay rates exists between both directions. Since the system cannot reside at half-stable equilibrium points due to the likely presence of noise, then we may infer from Proposition \ref{proposition:alternatingstability} in the Appendix that the eventual regime the system will reside in after the bifurcation, if such a regime exists, will lie along the direction of the regime shift. Concurrent to the phenomena of decay rate asymmetry is decreasing decay rate which can be used to verify that the system is losing stability. In order to apply the results of the previous section, we also require that $P(N)$ and $C(N)$ be smooth functions of $N$. Lag-1 autocorrelation is used to determine the level of critical slowing down while skewness is used to determine the level of asymmetry in the decay rates at the equilibrium point the system is residing in (Section \ref{sec:CriticalSlowingDownSignals} in the Appendix). It should be noted that it is possible for the time series to be autocorrelated or skewed without the system being close to a bifurcation point. Therefore, it is not the absolute value of these signals we are measuring but the relative changes of these signals we are measuring. By tuning an input parameter and measuring autocorrelation and skewness statistics of the fluctuations about the equilibrium point, the algorithm verifies that the changes in these signals conform to the results of Table \ref{table:CSD}. These signals are then used to determine whether to increase or decrease an input parameter to approach the bifurcation which will result in a regime shift in the desired direction. Finally, the parameter is tuned in the system until a regime shift occurs. The pseudocode of the algorithm is outlined here in the main text. Variables and constants used are defined in the pseudocode. A table listing the definitions of variables and constants used is also provided for reference (Table \ref{table:shapirowilk} in the Appendix). 

\begin{algorithm}[H]
\noindent\begin{minipage}{\textwidth}
\caption*{Engineering a regime shift}
\begin{algorithmic}[1]
\Procedure{EngineerRegimeShift}{}
\State $r \gets$ initialized bifurcation parameter of the system
\State $N' \gets$ observations of the state variable from the system with parameter $r$
\State $d \gets$ desired direction of regime shift ($+1$ for positive and $-1$ for negative direction)
\State $tol \gets$ tolerance level for defining a regime shift in the state variable
\State $window\_length \gets$ length of time windows used in the calculation of statistical signals
\State $increment \gets$ positive value to be added to or deducted from $r$
\State $N \gets$ \Call{BurnIn}{$N'$} \footnote{The function \textsc{BurnIn} is described in detail on the next page. It essentially truncates $N'$ from the front to allow the simulation time to reach equilibrium. }
\State $skewness$, $autocorr$ $\gets$ \Call{ComputeCSDSignals}{$N$, $window\_length$} \footnote{The function \textsc{ComputeCSDSignals} is described in detail on the next page. }
\State $r_+ \gets r + increment$
\State $r_- \gets r - increment$
\State $N_{r+}' \gets$ observations of the state variable from the system with parameter $r_+$
\State $N_{r-}' \gets$ observations of the state variable from the system with parameter $r_-$
\State $N_{r+}$ $\gets$ \Call{BurnIn}{$N_{r+}'$}
\State $N_{r-}$ $\gets$ \Call{BurnIn}{$N_{r-}'$}
\State $skewness_{r+}$, $autocorr_{r+} \gets$ \Call{ComputeCSDSignals}{$N_{r+}$, $window\_length$}
\State $skewness_{r-}$, $autocorr_{r-} \gets$ \Call{ComputeCSDSignals}{$N_{r-}$, $window\_length$}
\State $h_{a,r+} \gets$ one-tailed Welch's t-test of $\Expectation[autocorr_{r+}]>$ $\Expectation[autocorr]$\footnote{The symbols $h$ are Boolean variables of the acceptance of the alternative hypothesis stated. In this case, $h_{a,r+}$ is true if the alternative hypothesis $\Expectation[autocorr_{r+}]>$ $\Expectation[autocorr]$ is accepted and false if the hypothesis testing is inconclusive. } 
\State $h_{a,r-} \gets$ one-tailed Welch's t-test of $\Expectation[autocorr_{r-}]>\Expectation[autocorr]$
\State $h_{s,r+} \gets$ one-tailed Welch's t-test of $d\times \Expectation[skewness_{r+}]>d\times$ $\Expectation[skewness]$
\State $h_{s,r-} \gets$ one-tailed Welch's t-test of $d\times \Expectation[skewness_{r-}]>d\times$ $\Expectation[skewness]$
\If {$h_{a,r-}$ \aand $h_{a,r+}$ are \False} 
\Raise inconclusive test for increasing autocorrelation
\EndIf
\If {$h_{s,r-}$ \aand $h_{s,r+}$ are \False} 
\Raise inconclusive test for changing skewness
\EndIf
\algstore{engineer}
\end{algorithmic}
\end{minipage}
\end{algorithm}

\begin{algorithm}[H]
\noindent\begin{minipage}{\textwidth}
\caption*{Engineering a regime shift}
\begin{algorithmic}[1]
\algrestore{engineer}
\If {$h_{a,r-}$ \aand $h_{a,r+}$ are \True \oor \\ \hspace\algorithmicindent\hspace{0.35cm} $h_{s,r-}$ \aand $h_{s,r+}$ are \True \oor \\\hspace\algorithmicindent\hspace{0.35cm} $h_{a,r+}$ \aand $h_{s,r-}$ are \True \oor \\\hspace\algorithmicindent \hspace{0.35cm} $h_{a,r-}$ \aand $h_{s,r+}$ are \True}
\Raise increasing CSD signals in both parameter directions
\EndIf
\If {$h_{a,r+}$ \aand $h_{s,r+}$ are \True}
\State $tuning\_direction \gets 1$
\Print Increasing parameter will lead to regime shift in desired direction
\EndIf
\If {$h_{a,r-}$ \aand $h_{s,r-}$ are \True}
\State $tuning\_direction \gets -1$
\Print Decreasing parameter will lead to regime shift in desired direction
\EndIf
\State $prev\_autocorr \gets autocorr$
\State $prev\_N \gets N$
\Do
\State $r \gets r + tuning\_direction \times increment$
\State $N' \gets $ Observations of the state variable from the system with parameter $r$
\State $N \gets $ \Call{BurnIn}{$N'$}
\State $h_f \gets$ one-tailed Welch's t-test of $d\times\Expectation[N]>d\times\Expectation[prev\_N]+tol$
\If {$h_f$ is \True}
\Print Regime shift has occurred
\State \Return
\EndIf
\State $skewness$, $autocorr \gets$ \Call{ComputeCSDSignals}{$N$, $window\_length$}
\State $h_{a-} \gets$ one-tailed Welch's t-test of $\Expectation[autocorr]<\Expectation[prev\_autocorr]$
\If {$h_{a-}$ is \True}
\Raise Autocorrelation decreases when tuning parameter
\EndIf
\State $h_{a+} \gets$ one-tailed Welch's t-test of $\Expectation[autocorr]>\Expectation[prev\_autocorr]$
\If {$h_{a+}$ is \False}
\Raise Inconclusive test for increasing autocorrelation
\EndIf
\State $prev\_N \gets N$
\State $prev\_autocorr \gets autocorr$
\doWhile{$h_f$ is \false}
\EndProcedure
\Function{ComputeCSDSignals}{$N$, $window\_length$} \footnote{More details on the calculation and formulas used for computing of the CSD signals can be found in the Section \ref{sec:CriticalSlowingDownSignals} in the Appendix}.
\State Segment $N$ into non-overlapping time windows of length $window\_length$
\For{each $time\_window$ in $N$}
\State $skewness[\text{index of } time\_window] \gets$ skewness of $time\_window$
\State $autocorr[\text{index of } time\_window] \gets$ lag-$1$ autocorrelation of $time\_window$
\EndFor
\State \Return $skewness$, $autocorr$
\EndFunction
\Function{BurnIn}{$N'$}
\State Bin the observations $N'$ and obtain frequency counts for the bins
\State $b \gets$ interval of bin with highest frequency count
\State $N \gets$ $N'$ truncated from the front by removing observations up till the first observation in $b$ 
\State \Return $N$
\EndFunction
\end{algorithmic}
\end{minipage}
\end{algorithm}

For this algorithm, we have to choose an appropriate tolerance level (line 4 of the pseudocode) and a window length (line 5 of the pseudocode). The tolerance level is used for identifying when a regime shift has occurred (line 52 of the pseudocode). The tolerance level should be larger than the change in the equilibrium point at each iteration when the parameter is tuned, but small enough so that a regime shift that has occurred will not go undetected. The window length is the number of elements in each time window. Time windows are obtained by segmenting the time series of fluctuations. The non-overlapping segments or time windows are used in estimating the statistical signals and their statistical significance in the time series generated by a stationary ergodic process. As the number of time windows increases, the mean of the statistical signals across all time windows will approach a normal distribution by the central limit theorem. To obtain an accurate estimate of a statistical signal, the window length should be as large as possible without compromising the normality assumption of the hypothesis tests. When calculating the burn-in to obtain the fluctuations about the equilibrium point (line 63 of the pseudocode), it might also be necessary to detrend the time series if there are seasonal fluctuations present in the data. 

This algorithm requires that the system to be already close to a bifurcation for the results of the previous section to apply. In order to create such a scenario, a parameter $r$ can be tuned continuously until a steady increasing trend of critical slowing down signals is observed in the tuning direction of $r$. It should be noted that depending on $f(N)$, this method is not infallible as it is possible for an increasing trend to be deemed statistically significant without approaching a bifurcation (see Section \ref{sec:CSDwoBifur} in the Appendix). In this case, the parameter must be tuned and explored further, failing which, in the case where $f(N)$ contains more than one parameter, we may need to keep switching to other parameters until the desired regime shift is achieved. 

Because this algorithm relies on measuring statistical signals of critical slowing down, the efficacy of this algorithm to bring about desired regime shifts will depend on the nature of noise in the system. For example, noise that is biased against the trend of critical slowing down signals can mar the ability to determine if a parameter is a bifurcation parameter. Another potential complication is the statistical significance of skewness measurements. The skewness is harder to detect than the autocorrelation because the decay rates are roughly symmetric in both directions when perturbations are small. This problem can be resolved when the variance of fluctuations is large enough and a large number of observations are taken. The variance of fluctuations is itself a critical slowing down signal so that the significance of skewness observations becomes easier to determine closer to the bifurcation as the variance of fluctuations increases. 

\begin{figure*}[h!]
\centering
\centerline{\includegraphics[scale=0.5]{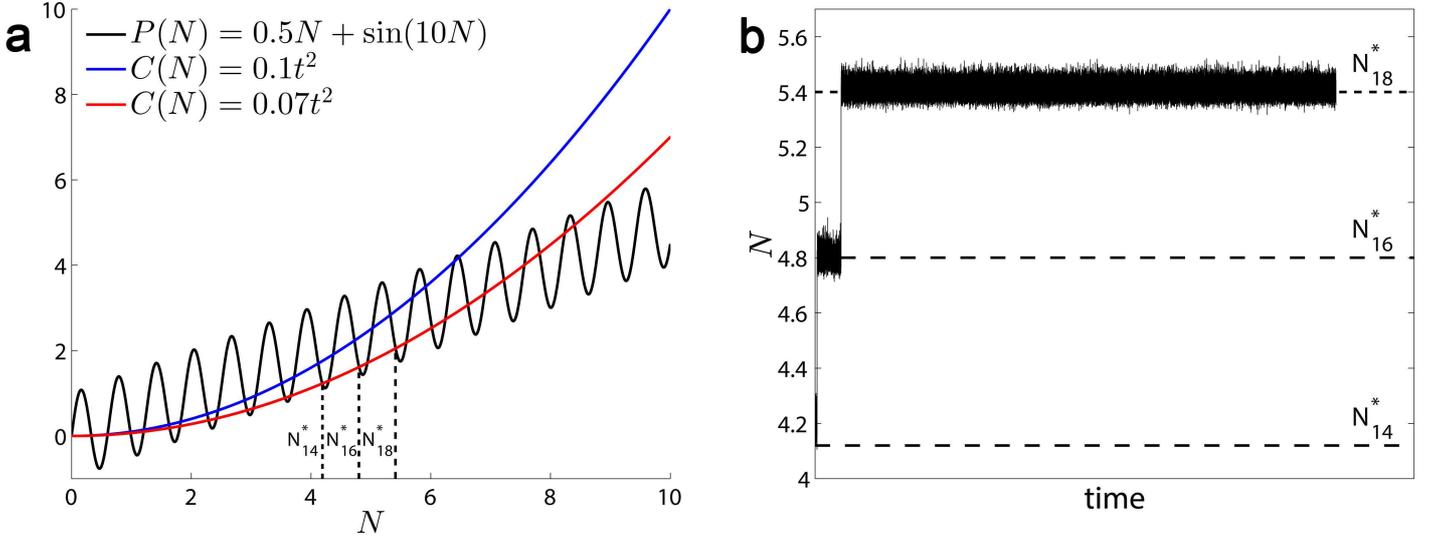}}
\caption{(a) The functions $P(N)$ and $C(N)$ given by Eq. \ref{eq:example} and \ref{eq:example2} are plotted against $N$ for $r=0.1$ and $r=0.07$. The stable equilibrium points $N^*_{14}$ and $N^*_{18}$ when $r=0.07$ are marked by the dashed lines. (b) The time series obtained by numerical integration of Eq. \ref{eq:example2} using the Euler-Maruyama method when $r=0.07$. Dashed lines mark the stable equilibrium points $N^*_{14}$, $N^*_{16}$ and $N^*_{18}$ when $r=0.07$. } 
\label{fig:fN_}
\end{figure*}

Here, we implement this algorithm in MATLAB on a one-dimensional dynamical system exhibiting multiple regimes: 
\begin{align} \label{eq:example}
dN &= \left[ P(N) - C(N) \right]dt + \sigma dW_t, \\ \label{eq:example2}
&= (0.5N + \sin(10N) - rN^2)dt + \sigma dW_t,
\end{align}
where $W_t$ is the standard Wiener process (Brownian motion), $r$ is a bifurcation parameter, $P(N)=N+\sin (N)$, and $C(N)=rN^2$. For the purposes of demonstration, we loosen the restriction that $P(N)$ is strictly increasing and non-negative. Plots of $P(N)$ and $C(N)$ can be seen in Fig. \ref{fig:fN_}(a). 

The Euler-Maruyama method was used to obtain numerical observations from Eq. \ref{eq:example2}. A step size of 0.001 was chosen for the numerical method and the starting system parameters are $N(0)=N^*_{14}|_{r=0.1}$, $r=0.1$, and $\sigma=0.08$. Here, $N^*_{i}$ refers to the $i$th equilibrium point in the system. A tolerance level of 0.2 and a window length of 10,000 data points are used. We want to engineer a regime shift in the positive direction from $N^*_{14}|_{r=0.1}$. The algorithm identified that a decrease in $r$ when $r=0.1$ would bring the system closer to a bifurcation in the desired direction (positive). After decreasing $r$ sequentially from $0.1$ to $r=0.07$ in decrements of $0.01$, a regime shift occurred bringing the system from $N^*_{14}|_{r=0.07}$ to a transient regime $N^*_{16}|_{r=0.07}$ before settling at $N^*_{18}|_{r=0.07}$ thereafter (Fig. \ref{fig:fN_}(b)). Although no local bifurcations occurred from $r=0.1$ to $r=0.07$ (Fig. \ref{fig:fN_}(a)) involving the equilibrium point the system was residing in, the weakening stability allowed the system to escape $N^*_{14}|_{r=0.07}$ due to the noise in the system. The bifurcation that would annihilate  $N^*_{14}$ and $N^*_{15}$ is a saddle node bifurcation (Fig. \ref{fig:bifurDiag} in the Appendix). Outcomes of hypothesis testings in the course of implementing the algorithm on this model can be found in Table \ref{table:hypothesistestings} in the Appendix. Furthermore, we also conducted normality tests on the distributions of the critical slowing down signals to ascertain that the window length used is appropriate (Fig. \ref{fig:qq} and Table \ref{table:hypothesistestings} in the Appendix). 

\section*{Discussion}
By generalizing Bettencourt et al.'s model of population dynamics, we have incorporated mechanisms for the stagnation of a population due to declining profitability in a city's industrial output. The stagnation occurs when the population settles at a stable equilibrium point. Here, in order to drive the city into a growth phase and prevent urban decline, a regime shift has to be engineered from this stable equilibrium point. To the best of our knowledge, we have not yet observed any algorithms in the literature for the systematic engineering of regime shifts  likely because regime shifts are mostly viewed negatively as undesirable events in the ecology literature. However, through the measurement of critical slowing down signals, we can tell if tuning a parameter will result in a loss of stability and eventually lead to a bifurcation, with the skewness giving us the direction where the resulting regime will lie relative to the present one. This result gives us confidence in the direction of the impending regime shift and is then used to present an algorithm based on the measurement of these statistical signals that is capable of bringing about a regime shift such that the city may recover from stagnation and continue growing if it is able to identify and invest in profitable industries. 

While the algorithm can determine if a parameter being tuned is a potential bifurcation parameter, accurate identification of the bifurcation parameters can shorten the time needed in implementing the algorithm in real life. If we can only obtain population change estimates every month, and thereafter need multi-year time windows to reliably detect CSD, using the algorithm to bring about a regime shift can take decades. Therefore, if possible, we would like to accurately identify the bifurcation parameter to avoid wasting years or even decades tuning parameters that do not eventually lead to the desired regime shift. We imagine the actions and policies involved can be more accurately identified through realistic agent-based models at the firm and household level, but the development of such models is beyond the scope of this paper.

The validation of the population dynamics model presented in this paper will need to depend on the detection of regimes and regime shifts in cities. Some of the criteria used for judging the presence of empirical regimes in the ecology literature have relied on controlled experiments \cite{Schroder}, something that is hard to replicate for cities. A simpler method is to detect critical slowing down signals preceding a large observed change in the population of a city. This requires higher frequency time series than what was historically available on publicly available census data. If, for example, the urban decline of Detroit indeed results from a regime shift to a regime with a lower population, then this transition happens on the order of decades. However, census population data for the second half of the 20th century was only collected every decade, which is too sparse for the calculation of the statistical signals of critical slowing down. With the onset of big data and social media, it should be possible to obtain higher frequency population estimates of a city without much effort relative to traditional census collection methods. It is not entirely clear at the moment whether a not the population dynamics of a real city may involve regimes and regime shifts. However, a regime shift in the US housing market was detected across multiple cities \cite{Tan,JPLTanEPJB}. This regime shift was associated with a large abrupt increase in the proportion of subprime mortgages issued in the United States prior to the subprime crisis. If regimes and regime shifts can exist within an economy, then it is reasonable to expect the population of a city to experience similar phenomena given the dependence of a city's population on its economy. 

In addition to verifying that real-world cities undergo regime shifts, the simplicity of the model considered here comes at the cost of the assumptions we make. It is not clear if this simple model is able to capture the essential economic mechanisms of urban decline and urban growth. Therefore, moving on from the simple urban population dynamics model, the next stage of research on engineering regime shifts in cities would involve validating the model and testing the algorithm on more realistic computer simulations of the population dynamics of a city. Such a simulation should be of a bottom-up nature since cities are dense spatial agglomerations of individuals competing for space and resources. Indeed, cities are examples of complex adaptive systems where autonomous individuals continuously adapt to and interact with other individuals and the environment, giving rise to complex emergent phenomena \cite{Batty1, Chen1}. Here, agent-based computer simulations are a natural candidate for the job as they are able to reproduce these emergent phenomena in addition to providing a realistic picture of the bottom-up processes driving the formation and evolution of a city \cite{Bonabeau1}. Presently, there exist agent-based models for the population dynamics of a city due to economic migration but these models do not simulate a functioning economy of the city \cite{Benenson1}. Conversely, there exist agent-based models of economies that do not simulate the population dynamics of a city \cite{Tesfatsion1}. An ideal agent-based model must integrate both approaches successfully before it can be used to investigate the engineering of regime shifts for the economic revitalization of a city. 

Due to the generality of the assumptions stated, the algorithm is easily applicable to dynamical models in other fields such as ecology where the presence of regimes and regime shifts are well established. Furthermore, an algorithm that can be used for the stabilization of an equilibrium point rather than engineering regime shifts could easily follow from the concepts covered in this paper. Therefore, we believe that the ideas discussed here constitutes one of many crucial first steps to realize greater control over the sometimes unpredictable nature of non-linear complex systems. 

\section*{Acknowledgments}
The author would like to thank Siew Ann Cheong for proposing the problem of engineering regime shifts and for reviewing the manuscript before submission. The author would also like to thank Youngho Chang for comments on the manuscript. 

\clearpage
\renewcommand\thefigure{A\arabic{figure}}
\renewcommand\theequation{A\arabic{equation}}
\renewcommand\thetable{A\arabic{table}}
\setcounter{figure}{0}

\part*{Appendix}
\section{Alternating stability in $f(N)$}

\begin{proposition} \label{proposition:alternatingstability}
The stability of the equilibrium points in the sequence $N^*$ always alternate between stable and unstable, not counting the equilibrium points that are half-stable. 
\end{proposition}

\textit{Proof.} Since $P(N)$ and $C(N)$ are continuous functions of $N$, $f(N)$ is also a continuous function of $N$. Let $N^*_i \in N^*$ be an equilibrium point such that $i<n$ and let $j=i+1$. Because $f(N)$ is continuous, then $f(N^*_{i+})$ must have the same sign as $f(N^*_{j-})$. Similarly, if $i>1$ and $k=i-1$, then $f(N^*_{i-})$ must have the same sign as $f(N^*_{k+})$. Hence, the result stated is obtained. \hfill$\square$ 

\section{Increasing critical slowing down signals without approaching a bifurcation} \label{sec:CSDwoBifur}
As an example of a situation where a continuous unidirectional tuning of a parameter causes an increasing trend of critical slowing down signals without approaching a bifurcation, consider the one-dimensional dynamical system 
\begin{align}
\dot{x} = c + r^2-x^2, 
\end{align}
where $c>0$ is a constant, $x$ is the state variable and $r$ is the parameter we are tuning. For any $r$ and $c$, there exists two equilibrium points $x^*_{\pm} = \pm \sqrt{c+r^2}$, with $x^*_+$ being stable and $x^*_-$ being unstable. By a translation $y=x-x^*$, we obtain $\dot{y} = g(y) = -y^2-2y\sqrt{c+r^2}$. Let $\varepsilon>0$ be the magnitude of a perturbation from $x^*$ within the basin of attraction of $x^*$. Hence, a perturbation $\pm \varepsilon$ decays as $g(\pm \varepsilon)$. We see that if $r$ is positive and is decreased continuously, then an increasing trend of critical slowing down signals would be detected as $|g(\pm\varepsilon)|$ decreases with decreasing $r$. However, this trend would reverse when $r<0$ such that the trend of critical slowing down signals is decreasing as $r$ is negative and is continually decreased. Hence, a bifurcation is never reached through a unidirectional tuning of a parameter even though an increasing trend of critical slowing down signals is initially observed. 

\section{Critical slowing down signals} \label{sec:CriticalSlowingDownSignals}
\subsection{Autocorrelation}
Autocorrelation measures the memory of a time series. As an equilibrium point loses stability, increasing memory in the time series of the state variable occurs because of a decreasing decay rate to equilibrium \cite{SchefferScience1}. Here, we calculate the lag-1 autocorrelation
\begin{align}
r = \frac{\sum_{i=2} (y_i - \bar{y})(y_{i-1}-\bar{y})}{{s_y}^2}
\end{align}
, where $y=\{y_1, y_2, y_3, \dots \}$ is a time series for which we are measuring the lag-1 autocorrelation $r$, $\bar{y}$ is the mean and ${s_y}$ is the standard deviation of the time series. 

\subsection{Skewness}
The skewness is also a critical slowing down signal and it measures how asymmetric a distribution is about its mean \cite{GuttalEcolLett1}. For some kinds of bifurcations, the loss of stability in one direction is greater than the other direction (Main text Section 3), causing the distribution of the state variable to be skewed towards the direction with the lower decay rate. The skewness of a time series $y$ is
\begin{align}
\gamma = \sum_{i=1} \left( \frac{y_i - \bar{y}}{s_y} \right)^3. 
\end{align}

\section{Engineering a regime shift on Eq. \ref{eq:example}}
In the main text, we engineered a regime shift in the positive direction on the following dynamical system
\begin{align} \label{eq:example}
dN = (N + \sin(N) - rN^2)dt + \sigma dW_t,
\end{align}
using the algorithm presented when $N(0)=N^*_{14}|_{r=0.1}$ and $r=0.1$. In Table \ref{table:hypothesistestings}, we provide detailed outcomes of the hypothesis testings in the course of implementing the algorithm. 

\begin{figure*}[h!]
\centering
\centerline{\includegraphics[scale=0.5]{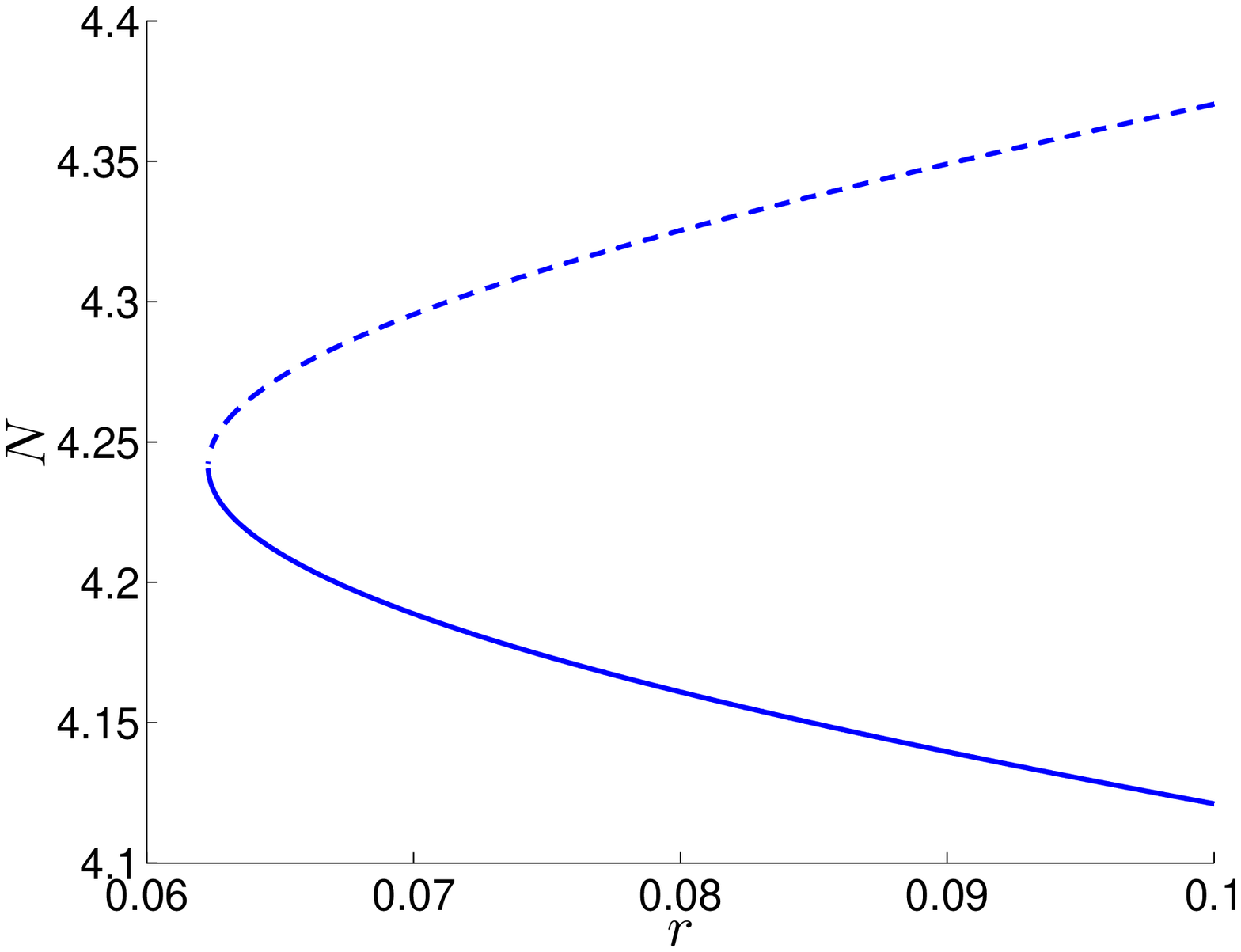}}
\caption{Bifurcation diagram of the system at $N^*_{14}$ as it approaches a saddle-node bifurcation from $r=0.1$ to $r=0.062$. The dashed line is the unstable equilibrium point $N^*_{15}$ and the solid line is the stable equilibrium point $N^*_{14}$. Both indices $14$ and $15$ refer to the same equilibrium point from $r=0.1$ to $r=0.062$ since there is no change in the number of equilibrium points before $N^*_{14}$ from $r=0.1$ to $r=0.062$. } 
\label{fig:bifurDiag} 
\end{figure*}

\begin{table}[h!] 
\caption{Outcomes of hypothesis testings in the course of implementing the algorithm to engineer a regime shift on Eq. \ref{eq:example} as described in the main text. }
\centering
\begin{tabular}{l | l l l l} \label{table:hypothesistestings}
 & $r=0.1$ & $r=0.09$ & $r=0.08$ & $r=0.07$ \\ \hline 
$h_s$ & True (1.2E-27) &  N/A & N/A & N/A \\ 
$h_{a,r+}$ & False (1) & N/A & N/A & N/A \\
$h_{a,r-}$ & True (9.2E-56) & N/A & N/A & N/A \\
$h_{s,r+}$ & False (1) & N/A & N/A & N/A \\ 
$h_{s,r-}$ & True(3.0E-4) & N/A & N/A & N/A  \\ 
$h_f$ & N/A & False (1) & False (1) & True (1)  \\ 
$h_{a-}$ & N/A & False (1) & False(1) & N/A \\
$h_{a+}$ & N/A & True (2.2E-61) & True (7.5E-118) & N/A
\end{tabular}
\end{table}

\subsection{Normality tests}
\begin{figure*}[h!]
\centering
\centerline{\includegraphics[scale=0.3]{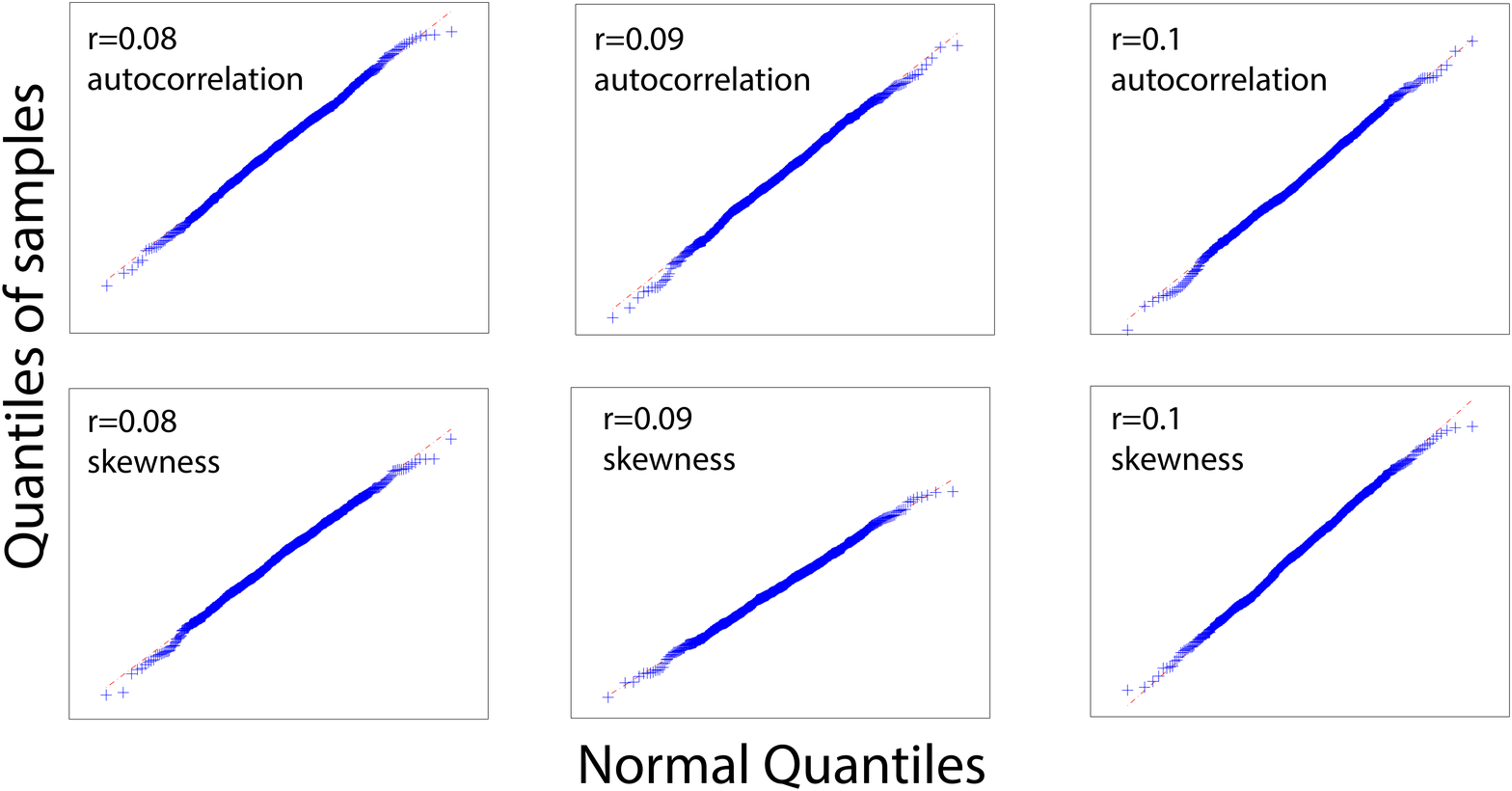}}
\caption{Q-Q plots of the autocorrelation and skewness distribution of time windows at the various values of $r$ when engineering the regime shift on Eq. \ref{eq:example}. } 
\label{fig:qq} 
\end{figure*}

To ascertain that the normality assumption is reasonable when using Welch's t-test in comparing the statistical signals at different values of $r$, we conducted a visual inspection of the Q-Q plots of these distributions against the normal distribution (Figure \ref{fig:qq}). We also conducted normality tests, namely Shapiro-Wilk tests at the 5\% significance level (Table \ref{table:shapirowilk}), and found that that the tests resulted in a failure to reject the null-hypothesis of normality for the distributions at the values of $r$ encountered. Hence, we conclude that the normality assumption is reasonable and that the window length chosen for the time windows is appropriate. 

\begin{table}[h!] 
\caption{$p$-values of the Shapiro-Wilk test of normality at the 5\% significance level conducted on the autocorrelation and skewness distribution of time windows. The $p$-values are calculated at the various values of $r$ when engineering the regime shift on Eq. \ref{eq:example}. }
\centering
\begin{tabular}{l | l l l l} \label{table:shapirowilk}
 & $r=0.1$ & $r=0.09$ & $r=0.08$ \\ \hline 
autocorrelation & 0.31 & 0.19 & 0.35  \\ 
skewness & 0.38 & 0.75 & 0.11 \\
\end{tabular}
\end{table}

\begin{table}[h!]
\caption{Definitions of variables and constants used in the pseudocode. Text in parentheses indicate values for constants used when engineering a regime shift on Eq. \ref{eq:example}.}
\label{table:pseudocode}
\centering
\begin{tabularx}{\textwidth}{l X}
\toprule
\tabhead{Variable/Constant} & \tabhead{Definition} \\
\midrule
$r$ & Proposed bifurcation parameter of the system (intialized at 0.1) \\
$r+$ & $r+increment$ \\
$r-$ & $r-increment$ \\
$N'$ & Time series of the observed state variable with the system at $r$\\
$N'_{r+}$ & Time series of the observed state variable with the system at $r+$\\
$N'_{r-}$ & Time series of the observed state variable with the system at $r-$\\
$N$ & $N'$ after burn in where $N'$ is truncated from the front to allow the system to reach equilibrium \\
$N_{r+}$ & $N'_{r+}$ after burn in where $N'_{r+}$ is truncated from the front to allow the system to reach equilibrium \\
$N_{r-}$ & $N'_{r-}$ after burn in where $N'_{r-}$ is truncated from the front to allow the system to reach equilibrium \\
$d$ & The desired direction of the regime shift, +1 for positive and -1 for negative direction (+1) \\
$tol$ & Tolerance level for defining a regime shift in the state variable (0.2) \\
$window\_length$ & Length of time windows used in the calculation of statistical signals (10,000) \\
$increment$ & A positive value to be added to or deducted from $r$ (0.01) \\
$skewness$ & Array of skewness values corresponding to each time window with the system at $r$\\
$skewness_{r+}$ & Array of skewness values corresponding to each time window with the system at $r+$\\
$skewness_{r-}$ & Array of skewness values corresponding to each time window with the system at $r-$ \\
$autocorr$ & Array of lag-1 autocorrelation values corresponding to each time window with the system at $r$\\
$autocorr_{r+}$ & Array of lag-1 autocorrelation values corresponding to each time window with the system at $r+$\\
$autocorr_{r-}$ & Array of lag-1 autocorrelation values corresponding to each time window with the system at $r-$\\
$h_{a,r+}$ & Result of one-tailed Welch's t-test of $d\times \Expectation [autocorr_{r+}]>\Expectation [autocorr]$ \\
$h_{a,r-}$ & Result of one-tailed Welch's t-test of $d\times \Expectation [autocorr_{r-}]>\Expectation [autocorr]$ \\
$h_{s,r+}$ & Result of one-tailed Welch's t-test of $d\times \Expectation [skewness_{r+}]>\Expectation [skewness]$ \\
$h_{s,r-}$ & Result of one-tailed Welch's t-test of $d\times \Expectation [skewness_{r-}]>\Expectation [skewness]$ \\
$h_f$ & Result of one-tailed Welch's t-test of $d\times \Expectation [N]> d\times \Expectation [perv\_N] + tol$ \\
$h_{a+}$ & Result of one-tailed Welch's t-test of $\Expectation [autocorr]>\Expectation [prev\_autocorr]$ \\
$h_{a-}$ & Result of one-tailed Welch's t-test of $\Expectation [autocorr]<\Expectation [prev\_autocorr]$ \\
$tuning\_direction$ & Direction to tune parameter; +1 to increase parameter and -1 to decrease parameter (determined by algorithm to be -1) \\
$prev\_autocorr$ & Array of autocorrelation values in the previous iteration when tuning the bifurcation parameter \\
$prev\_N$ & Truncated time series in the previous iteration when tuning the bifurcation parameter \\
$b$ & Interval of bin with the highest frequency count observed after binning the time series $N'$, $N'_{r+}$ or $N'_{r-}$ in the function \textsc{BurnIn} \\
\bottomrule
\end{tabularx}
\end{table}

\end{document}